\documentclass[sigconf,nonacm]{acmart}
\newcommand{\code}[1]{{\small\texttt{#1}}}
\AtBeginDocument{%
  }

\setcopyright{acmlicensed}
\copyrightyear{2018}
\acmYear{2018}
\acmDOI{XXXXXXX.XXXXXXX}
\acmConference[IUI '27]{ACM International Conference on Intelligent User Interfaces}{February 8--11, 2027}{Helsinki, Finland}

\acmISBN{978-1-4503-XXXX-X/2018/06}

\usepackage{graphicx}
\begin{document}

\title{Linking Event-sequence Data Claims to Analytic Provenance in Chatbots: Technique and Evaluation}


\author{Po-Ming Law}
\affiliation{%
  \institution{Adaptive Machines, Inc}
  \city{San Francisco}
  \state{California}
  \country{USA}}
\email{terrance@adaptivemachines.ai}

\author{Jianheng Ouyang}
\authornote{Both authors contributed equally to this research.}
\author{Nguyen-Truong Thinh}
\authornotemark[1]
\affiliation{%
  \institution{The Hong Kong University of Science and Technology}
  \city{Hong Kong}
  \country{China}}
\email{jouyangae@connect.ust.hk}
\email{truongthinh.nguyen03@gmail.com}

\author{Weizhi Li}
\affiliation{%
  \institution{Adaptive Machines, Inc}
  \city{San Francisco}
  \state{California}
  \country{USA}}
\email{weizhi@adaptivemachines.ai}

\author{Arpit Narechania}
\affiliation{%
  \institution{The Hong Kong University of Science and Technology}
  \city{Hong Kong}
  \country{China}}
\email{arpit@ust.hk}

\renewcommand{\shortauthors}{Law et al.}

\begin{abstract}
Chatbots are increasingly used to analyze temporal event sequences of which website clickstreams are among the most common. Explanations are essential for understanding claims about event-sequence data. However, existing systems often require programming expertise to interpret these explanations and fail to explain the data processes underlying the data claims. We introduce a technique for linking a data claim to its analytic provenance (i.e., the analytic steps and results that produced it). It has two components: a reference table that records analytic steps and results in addressable cells, and a citation format that binds each claim to one or more cells. To operationalize this technique for clickstream analysis, we curated a set of analytic operations for clickstream analysis, organized them through a pivot-table abstraction, and developed tools that an AI agent can invoke to create a reference table and assemble a natural language response with cited data claims. To evaluate how citations affect users' assessment of data claims, we conducted two crowdsourced experiments with 90 participants each. The experiments showed that participants perceived correct data claims to be more accurate with citations when compared with no explanations, and that they perceived incorrect data claims to be less accurate with citations when compared with code and no explanations. \textbf{Supplementary materials:} https://prototypes-production-efcc.up.railway.app
\end{abstract}


\received{20 February 2007}
\received[revised]{12 March 2009}
\received[accepted]{5 June 2009}

\maketitle

\section{Introduction}

Event-sequence data are extensively collected across diverse domains, from healthcare~\cite{WRIGHT201573} to marketing~\cite{10.1145/846183.846188}. Clickstreams are among the most widely collected forms of behavioral data~\cite{Sakalauskas2024}. Any site or application instrumented with a web analytics platform produces them continuously. While many consumer chatbots and AI agents can automatically analyze this data, understanding the natural language responses could be challenging.

\begin{figure*}[t]
    \centering
    \includegraphics[width=\textwidth]{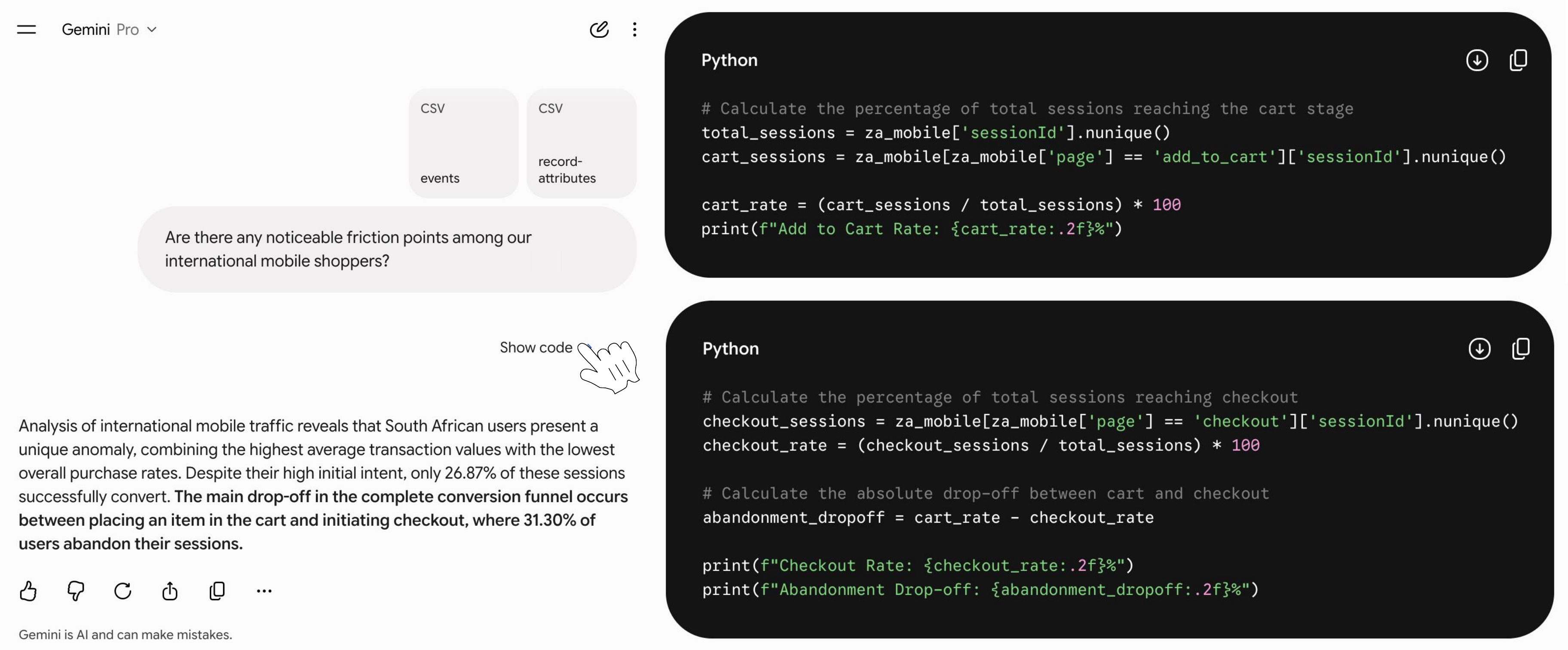}
    \caption{The prompt and response between a user and the Gemini application. This example comes from an actual chat session but has been simplified for clarity. The user asks, \textit{``Are there any noticeable friction points among our international mobile shoppers?''} The chatbot writes Python code and returns a natural-language response. Being curious about the analytic steps, the user clicks on ``Show code'' to reveal the generated Python code on the right.}
    \Description{A screenshot of a Gemini chatbot interface alongside two Python code snippets. On the left, a user uploads clickstream data and asks about friction points among international mobile shoppers. The chatbot identifies South African mobile shoppers and reports a 31.30 percent drop-off between adding an item to the cart and initiating checkout. A hand cursor points to the Show code option. On the right, two dark code panels show Python calculations: the first computes the percentage of sessions reaching the cart stage, and the second computes the percentage reaching checkout and subtracts the two percentages to obtain the reported drop-off.}
    \label{teaser}
\end{figure*}

Consider a product manager who uploads some clickstream data to a consumer chatbot such as Gemini~\cite{GeminiApp} and asks, \textit{``Are there any noticeable friction points among our international mobile shoppers?''} (Figure~\ref{teaser}). The dataset consists of a set of events. Each event records a session id, a timestamp, and the page visited. These events can be grouped by the session id and ordered by the timestamp to reconstruct a session (each visitor’s sequence of interactions). Each session carries attributes such as device type and country.

The chatbot writes Python, runs it, and reports that South Africa stands out: \textit{``...The main drop-off in the complete conversion funnel occurs between placing an item in the cart and initiating checkout, where 31.30\% of users abandon their sessions.''}

The sentence invites two questions. Which events make up the \textit{``complete conversion funnel''}? And 31.30\% of whom---of all South African mobile visitors, or only of the narrower cohort who placed an item in their cart? The product manager clicks ``Show code'' (Figure~\ref{teaser} left) to inspect the generated code. In Figure~\ref{teaser} right, the top code snippet computes the percentage of South African mobile users who added items to their shopping cart (67.3\% users did that). The bottom snippet then computes the percentage of users who reached the checkout stage (36\% users) and the percentage of users who abandoned their carts (67.3\% - 36\% = 31.3\%). Reassembling a claim from computations spread across snippets is laborious even for programmers, and this product manager does not program at all.

There are three considerations when designing an explanation technique that answers the product manager's questions. First, the explanations should not require users to read code. Existing systems that decompose AI-generated analysis into editable plans~\cite{Kazemitabaar2024} or interactive visualizations~\cite{WaitGPT2024} often target users proficient in programming.

Second, the explanations should provide the analytic provenance (i.e., analytic steps and results) behind the data claims (e.g., \textit{``31.30\%''}) within a natural language response. While some techniques offer stepwise explanations for AI-generated visualizations~\cite{wu2026stepmind}, they do not support linking text-based data claims to computational steps.

Third, the explanation technique should support event-sequence data. Existing tools like Aletheia~\cite{Fu2024Aletheia} provide citations for tabular data claims but cannot generalize to event sequences. These tools are often built on low-level analytic task taxonomies for tabular data (e.g.,~\cite{10.1109/INFOVIS.2005.24}) while event sequences involve temporal patterns, funnels, and sequential behaviors and have their own task frameworks~\cite{zinat2024multi, du2016coping, Guo2021SurveyEventSeq}.

To address these limitations, we ask, \textit{``How can we link an event-sequence data claim in a natural language response to the analytic steps and results that produced it, without requiring users to read code?''} We introduce a technique for linking event-sequence data claims to their analytic provenance. The technique has two components, one for representing provenance and one for linking claims to it. For representing provenance, we introduce a reference table. Each cell in this reference table records a sequence of analytic steps together with the results the steps produced. For linking claims to provenance, we introduce a citation format. Using this format, the chatbot wraps each data claim in its response in a tag referencing the cell elements that claim rests on.

\begin{figure}[h]
    \centering
    \includegraphics[width=0.4\textwidth]{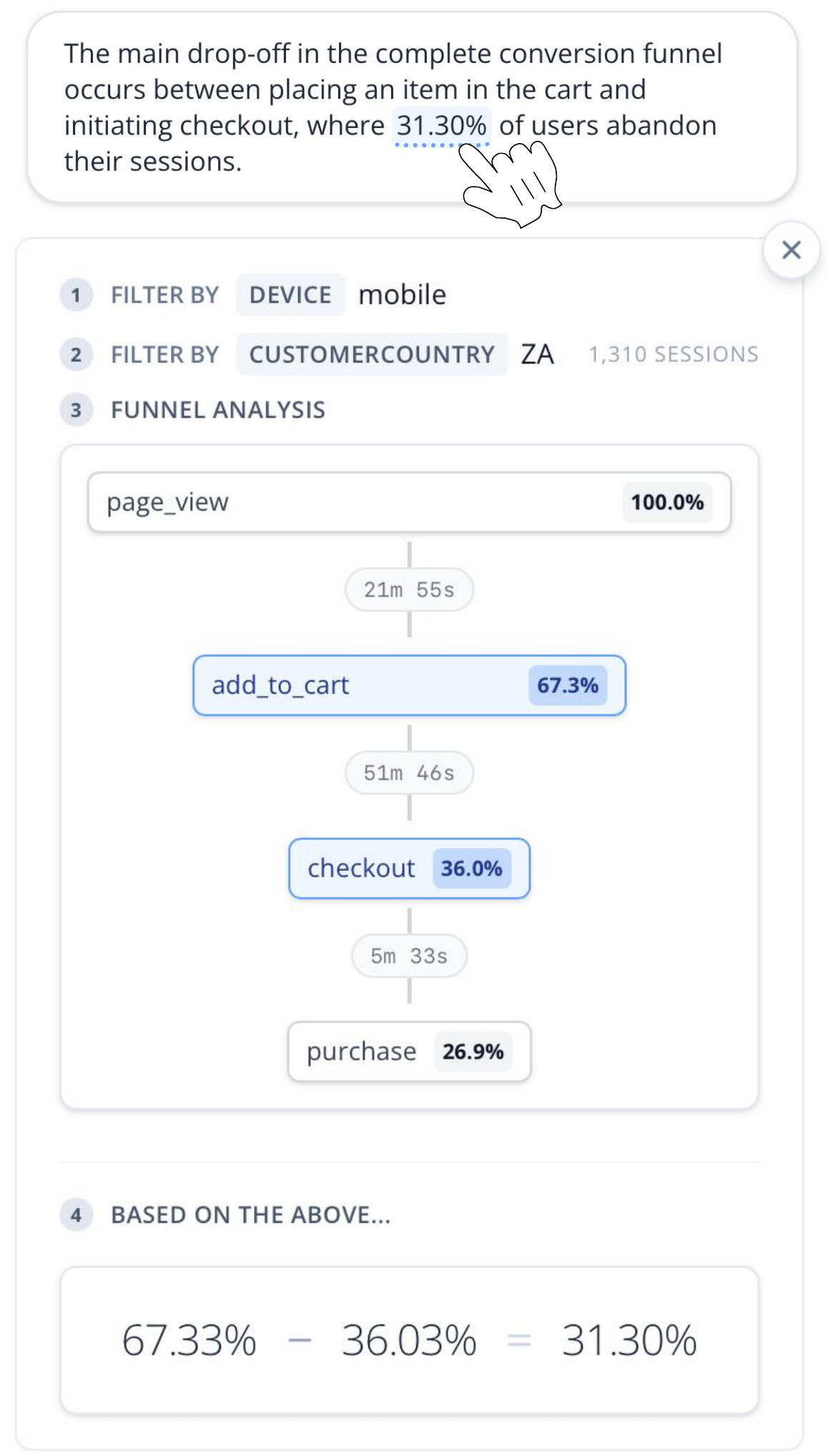}
    \caption{Using our technique, a user can hover over a data claim to see the analytic steps and results. Here, hovering over \textit{``31.30\%''} reveals that the system filters to sessions on mobile devices and from South Africa. The system then conducts a funnel analysis and finds that 67.3\% of all South African mobile users added items to their carts and 36\% checked out. 31.3\% is the difference between 67.3\% and 36\%.}
    \Description{A chatbot response with the value 31.30 percent highlighted and a floating explanation panel beneath it. The panel contains four numbered analytic steps. The first filters sessions by mobile device, and the second filters sessions by South Africa, leaving 1,310 sessions. The third shows a vertical conversion funnel with four stages: page view at 100 percent, add to cart at 67.3 percent, checkout at 36.0 percent, and purchase at 26.9 percent. The add-to-cart and checkout stages are outlined in blue. Durations between funnel stages appear along the connecting lines. The fourth step displays the calculation 67.33 percent minus 36.03 percent equals 31.30 percent.}
    \label{hover}
\end{figure}

Figure~\ref{hover} shows the effect. The product manager hovers over \textit{``31.30\%''} and a floating panel displays the analytic steps and results. In the explanation, the system first filters the sessions to focus on those using a mobile device from South Africa. It then evaluates a funnel with a page view event, an add to cart event, a checkout, and finally a purchase. This funnel clarifies that \textit{``complete conversion funnel''} refers to this sequence of four events. The funnel highlights that 67.3\% of all South African mobile users added items to their carts and 36\% of South African mobile users checked out. 31.3\% is the difference between 67.3\% and 36\%, indicating that 31.3\% of all South African mobile users abandoned their carts without checking out. A single hover addresses both of the product manager's questions without reading code.

To operationalize the technique, we curated a set of analytic operations for clickstream analysis based on event-sequence literature~\cite{zinat2024multi, du2016coping, Guo2021SurveyEventSeq} and expert consultations. We organized them through a pivot-table abstraction~\cite{Stolte2008Polaris} and developed a set of analytical functions that an AI agent can invoke to create a reference table with analytic provenance and link data claims in its response to the provenance. Following the terminology used in the AI agent community, we call these functions ``tools''~\cite{3666122.3669119, yao2022react}. These tools support operations such as selecting sessions by device or country, filtering sessions based on ordered event patterns, segmenting sequences between events, mining frequent patterns, and performing funnel analysis.

\begin{figure}[h]
    \centering
    \includegraphics[width=\linewidth]{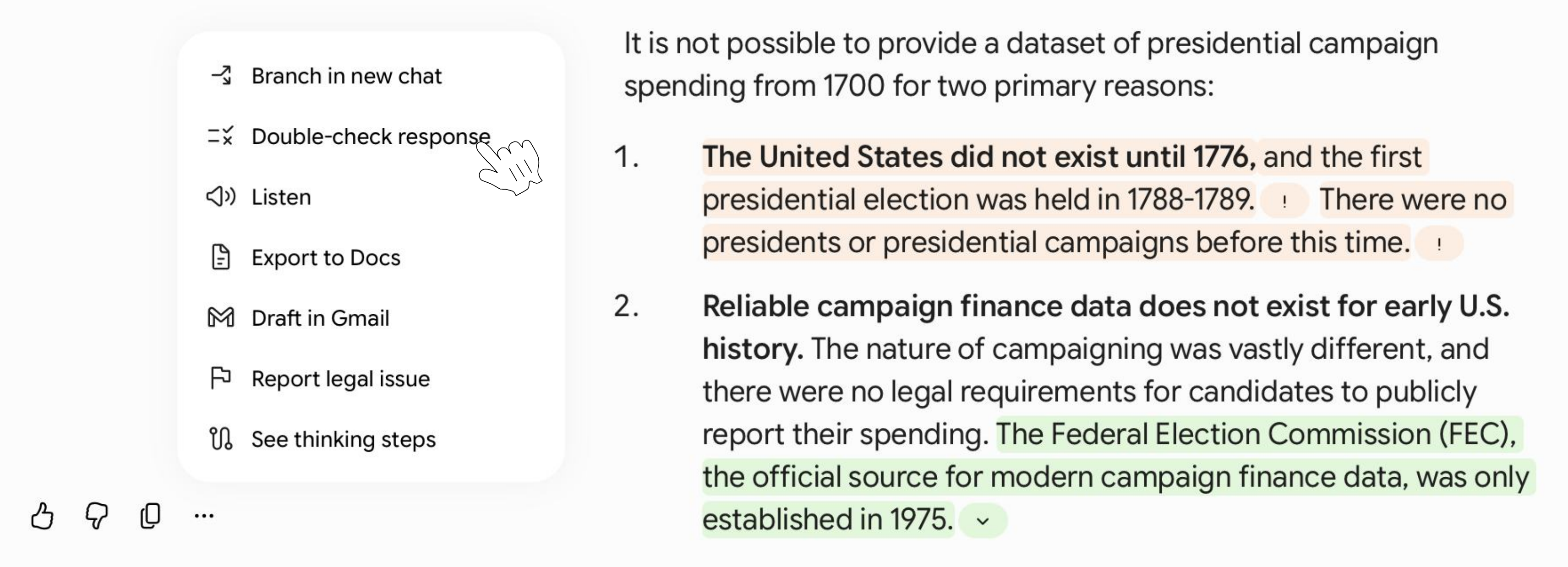}
    \caption{In the Gemini web application, users can select \textit{``Double-check response''} for certain responses. The system will highlight dubious claims in red and accurate claims in green. Users can click on a highlighted claim to check the sources.}
    \Description{Two screenshots from the Gemini web application. The left screenshot shows a dropdown menu containing the Double-check response option. The right screenshot shows a response after fact-checking, with portions of its text highlighted in different colors. A statement concerning presidential campaign finance data is highlighted in red, while a statement referencing the Federal Election Commission is highlighted in green. The highlighted passages indicate claims for which users can inspect supporting or conflicting web sources.}
    \label{double-check}
\end{figure}

We demonstrate the technique in two applications: automated insight generation (Figure~\ref{prototypes}a) and post-hoc fact-checking (Figure~\ref{prototypes}b). In the prototype for automated insight generation, users ask analytic questions and the chatbot answers with cited data claims. Many chatbots offer to fact-check a natural language response (Figure~\ref{double-check}), but they target facts checkable against the web. In our prototype for post-hoc fact-checking, users enter event-sequence data claims. The system then verifies the claims against the data and adds inline citations.

An explanation, however, is not automatically an improvement. Explanations can cause users to over-rely on AI output even when the output is wrong~\cite{Kim2025Fostering}. Hence, we ask, \textit{``How do citations of analytic provenance shape users' assessment of data claims?''} To address this question, we conducted two between-subject crowdsourced experiments corresponding to the two applications. In each experiment, 90 participants reviewed a series of data claims with no explanations, code snippets, or citations.

Experiment 1 examined automated insight generation in a setting where all presented claims were correct. We found that when a chatbot produced perfectly accurate insights, participants tended to perceive correct data claims with citations to be more accurate compared to no explanations. Experiment 2 examined post-hoc fact-checking in a setting where some claims contained controlled errors. When a chatbot occasionally produced errors, participants appeared to perceive incorrect data claims with citations to be less accurate compared to both code and no explanations.

Together, the experiments reveal an asymmetry: increasing perceived accuracy of correct claims while reducing perceived accuracy of incorrect ones. A central goal of explaining event-sequence data claims is to promote appropriate reliance: users trust AI-generated analyses when they are correct and reject them when they are not~\cite{Lee2004TrustAutomation}. This asymmetry is consistent with the pattern that appropriate reliance requires.

Our primary contributions include:

\begin{enumerate}
\item A technique for linking event-sequence data claims to their analytic provenance, comprising a reference-table representation of analytic steps and results and a citation format that binds claims within a natural-language response to cells of that table.
\item An operationalization of the technique for clickstream analysis: we curated nine analytic operations, organized them through a pivot-table abstraction and developed tools that AI agents can call to create a reference table of analytic provenance. 
\item Two prototypes demonstrating the technique for automated insight generation and for post-hoc fact-checking.
\item Findings from two crowdsourced experiments that compare how citations, code, and no explanations affect users' assessments of data claims within a chatbot’s responses.
\end{enumerate}

\section{Background and Related Work}

This paper draws on common terminology in the engineering of LLM-based agent systems, including agent and tool. Because these terms are used inconsistently across research and practice, we first define the terminology in this paper.

Our technique uses a set of tools that an agent can invoke to perform analytic operations on event sequence data. Hence, we review the operations commonly employed by systems for analyzing temporal event sequences. We further build on work that explains and fact-checks AI-generated analyses, and on approaches for evaluating such explanations.

\subsection{Terminology}

\subsubsection{LLM}

The AI agent engineering community often treats an LLM as a model endpoint: a stateless service that maps an input prompt to a text completion. 

\subsubsection{Agent}

An agent consists of a ``harness'' (i.e., infrastructure) built around the LLM. The harness includes system instructions that govern the LLM's behaviors and a set of ``tools'' (i.e., functions) that the LLM can invoke. With the harness, the agent calls tools to obtain results, feeds the results back to the LLM, which calls the tools in a control loop until the LLM produces a final answer~\cite{yao2022react}.

\subsubsection{Tool}

Tools are functions that an agent can invoke. Some tools call external services or APIs. For instance, a web-search tool can send a query to a search engine and return the results. When a tool integrates with such external services, the interface is often standardized through the Model Context Protocol (MCP). Other tools perform purely local computation and require no external API calls. For example, a tool can be a simple function that takes two numbers and returns their sum.

Tools also differ in whether they have side effects. For example, the tool that calculates the sum of two numbers does not have side effects. Tools that alter some external state when invoked have side effects. Consider a coding agent that operates within a sandbox: it may call a tool to write a new file to the sandbox and commit the changes to a GitHub repository.

\subsubsection{Chatbot}

A chatbot consists of a user interface built around an agent.

\subsubsection{Data Claim} 

We use data statement to refer to a sentence generated by an LLM based on the results of a data analysis. A data claim is a factual span within that statement whose meaning is supported by one or more analysis results. A data claim may report an exact value. For example, in \textit{``67.3\% of sessions involve adding an item to the cart,''} the claim \textit{``67.3\%''} is derived from a funnel analysis. A data claim may also express a qualitative characterization of magnitude or frequency. For example, in \textit{``A substantial proportion of sessions involve adding an item to the cart,''} the claim \textit{``a substantial proportion''} is grounded in the 67.3\% obtained from the same funnel analysis.

Our technique provides an agent with tools for performing analytic operations on temporal event sequences. We group these operations into three pivot-table operations~\cite{Stolte2008Polaris}: row operations, column operations, and filter operations. These tools do not make external API calls but have side effects: the agent operates within a workspace containing a reference table, and each tool invocation can add to or update that table. A chatbot built around this agent can associate event-sequence data claims in a natural-language response with entries in the reference table. When a user interacts with a cited claim, the interface displays the corresponding analytical steps and results (as in Figure~\ref{hover}).

\subsection{Analytic Operations in Event Sequence Analysis}

Researchers have developed visual analytic systems for analyzing event sequences in different application domains. For instance, Coreflow~\cite{liu2017coreflow} enables marketers to explore clickstreams and understand customer journeys. EventFlow~\cite{Monroe2013Temporal} supports investigating electronic health records while MAQUI~\cite{Law2018MAQUI} facilitates the exploration of emergency department workflow. These systems have been designed for different analytic purposes including cohort comparison~\cite{Krause_2016, Malik_2015}, pattern mining~\cite{Liu_2017, Vrotsou_2019}, and outcome prediction~\cite{Guo_2019, Kwon_2019}.

Several efforts have surveyed these systems and identified the analytic operations involved in event-sequence analysis. Notably, Du et al.~\cite{du2016coping} describe 15 strategies for reducing data volume and pattern variety during event-sequence analysis. Guo et al.~\cite{Guo2021SurveyEventSeq} categorized event-sequence analysis systems by five high-level analytical tasks. Zinat et al.~\cite{zinat2024multi} developed a hierarchical framework for describing event sequence analysis tasks. Across these accounts, recurring operations include filtering event sequences (e.g., extracting clickstream sessions from South Africa), segmenting sequences (e.g., focusing only on the events that happen after checkout), and mining frequent patterns.

Our technique exposes a set of tools that an agent can invoke to perform analytic operations on event sequences. These operations are grounded in the existing literature and feedback from an expert event sequence analyst.

\subsection{Explaining AI-Generated Data Claims}

Explanations of AI-generated answers take various forms depending on the application. They may aim to improve transparency~\cite{Liao2023AITransparency}, support comprehension~\cite{Kasneci2023ChatGPTEducation, Aydin2024LLMPatientEd}, or help users verify correctness~\cite{WaitGPT2024, Kazemitabaar2024}. When answering factual inquiries (e.g., who is the president of the United States), AI systems may provide supporting details~\cite{Lee2024OneVsMany, Xu2023CriticalEvaluation} or citations to external sources~\cite{Kim2024Dissatisfaction, Liu2023Verifiability} alongside the answer.

For data claims, Kazemitabaar et al.~\cite{Kazemitabaar2024} explored breaking an analysis into editable assumptions, plan, and code. Xie et al.~\cite{WaitGPT2024}, on the other hand, improved the readability of the generated code using interactive visualizations. Both explanation approaches target users proficient in programming. They also do not map a data claim within a natural language response to the corresponding analytic steps. StepMIND provides multimodal, stepwise explanations for AI-generated visualizations~\cite{wu2026stepmind}. However, it does not link individual claims in a textual response to their analytic provenance and does not address event-sequence analysis.

Besides research systems, consumer chatbots (e.g.,~\cite{GeminiApp, ChatGPT, ClaudeAI}) often present Python code to explain AI-generated data claims. As we discussed in the introduction, this approach presents two challenges: 1) it is difficult to trace which computations in the code support a data claim, and 2) understanding the code could be out of reach for users who lack programming expertise. We propose a technique for citing event-sequence data claims in a chatbot's interface and displaying the analytic provenance when users hover over a claim. Our proposed explanations do not require reading code.

\subsection{Fact-Checking AI-Generated Data Claims}

Automated fact checking has traditionally focused on verifying factual claims (e.g., Donald Trump is the 47th US president) that are not produced through data analysis (e.g., analyzing a CSV file). The veracity of factual claims can be automatically checked against knowledge bases such as Wikipedia pages~\cite{VitaminC} and scientific articles~\cite{MultiVerS}. For data claims, verification relies on inferences from data. Previous efforts translate claims into SQL queries to validate them against relational databases~\cite{AggChecker2019} or process chart images to check data statements visually~\cite{Akhtar2023ChartFactCheck}.

Consumer chatbots such as the Gemini application enable users to double-check factual claims (see Figure~\ref{double-check}). Dubious facts are then highlighted in red and correct facts in green. Users can click on a highlighted fact to check the sources. Inspired by this use case, we developed an application in which users enter event-sequence data claims and push a button to fact-check the claims (Figure~\ref{prototypes}b). The system cites the claims and users can hover over a claim to see the analytic steps. Closest to our work is Aletheia, which uses an LLM pipeline to parse data claims into structured fact specifications and executes those specifications against a reference dataset~\cite{Fu2024Aletheia}. Aletheia targets tabular data, however, and does
not support verifying event-sequence data claims.

\subsection{Evaluating Explanations for AI-Generated Data Claims}

Prior research has evaluated various types of explanations for AI-generated analyses. Examples include communicating accuracy information~\cite{He2023StatedAccuracy, Yin2019AccuracyTrust}, explanations of outputs~\cite{Bansal2021ComplementaryTeam, Schemmer2023AoR}, onboarding materials~\cite{Lai2020Chicago, Cai2021Onboarding}, and debugging interventions~\cite{He2024ToErr}.  Such evaluations often ask whether an explanation fosters appropriate reliance: broadly, users trust AI-generated analyses when they are correct and reject them when they are not~\cite{Lee2004TrustAutomation}.

Our crowdsourced experiments found that the citations improved perceived accuracy of correct event-sequence data claims compared to no explanations and reduced perceived accuracy of incorrect claims compared to both code and no explanations. Most relevant to our work, Kim et al.~\cite{Kim2025Fostering} found that clickable sources in LLM responses increase appropriate reliance on correct answers and reduce overreliance on incorrect ones, especially when paired with inconsistency highlighting. Their study examined citations to webpages for general factual questions. We instead studied the effect of showing analytic steps and results on users' assessment of data claims during event sequence analysis.

\section{Linking Event-Sequence Data Claims to Analytic Provenance}

Consider the statement from the introduction: \textit{``the main drop-off in the complete conversion funnel occurs between placing an item in the cart and initiating checkout, where 31.3\% of users abandon their sessions''}. Calculating the abandonment rate of 31.3\% involves 1) filtering the sessions by mobile devices, 2) filtering to those from South Africa, 3) performing a funnel analysis, and 4) computing the drop-off from adding to cart to checkout (see the four steps in Figure~\ref{hover}).

Our goal is to retain these analytic steps and results and associate them with the corresponding claim in the chatbot response. When users hover over \textit{``31.3\%,''} the chatbot can then reveal which sessions were analyzed, how the funnel was defined, and how the value was calculated.

In doing so, we designed a technique to store the analytic steps and results in a reference table. In this section, we will discuss how an AI agent can invoke tools to construct this reference table (Sec.~\ref{sec:table}). We will then showcase two use cases: automatic insight generation (Sec.~\ref{sec:insight}) and post-hoc fact-checking (Sec.~\ref{sec:fact-check}).

\subsection{Constructing Reference Table}
\label{sec:table}

Based on the survey of event sequence analytic operations~\cite{zinat2024multi, du2016coping, Guo2021SurveyEventSeq} and feedback from an expert event sequence analyst, we curated a set of analytic operations for analyzing clickstream data. We organized these analytic operations through a pivot-table abstraction~\cite{Stolte2008Polaris} and developed tools that an AI agent invokes to perform operations on event sequences. Using these tools, the agent constructs a reference table in which each cell records the analytic steps and results.

\subsubsection{Analytic Operations}
\label{sec:operations}

Zinat et al.~\cite{zinat2024multi} identify a wide range of event-sequence analytic operations, highlighting that the relevance of these operations varies across application domains. Hence, as opposed to creating a system that supports all possible analytic operations, we focused on the operations in a specific application domain while ensuring that our framework is extensible: to support new applications, we should be able to easily plug in new analytic operations into our framework. We will discuss the extensibility of our technique in Section~\ref{sec:extend}.

To identify analytic operations in a specific application domain, we collaborated closely with an expert analyst who has more than ten years of experience in event sequence analysis. At the time of our collaboration, the analyst worked at an e-commerce platform company and regularly analyzed clickstream data. We met weekly over two months. Earlier meetings focused on understanding the analyst's data and workflow. Subsequent meetings focused on refining the analytic operations in our framework. During these meetings, we presented a set of operations, obtained feedback, and refined the set until no additional changes were proposed.

Because the collaborator's data are proprietary, we cannot discuss their data in this paper. Their data have a similar structure to a Kaggle e-commerce clickstream dataset~\cite{ElHusseini2024EcommerceClickstream}. We will use this dataset as a running example.

In this Kaggle dataset, an event represents a user interaction and can be defined as a set of attribute-value pairs: $E = \{A_1 = V_1, A_2 = V_2, \ldots, A_j = V_j\}$. It has a session id, a timestamp, and attributes such as page (with values including \code{page\_view}, \code{add\_to\_cart}, and \code{purchase}). We call the attributes associated with an event event attributes.

An event sequence (or a session) is an ordered list of events: $S = [E_1, E_2, \ldots, E_k]$. It is constructed by grouping events with the same session id and ordering them by timestamp. Each event sequence has attributes describing the properties of the sequence. We call the attributes associated with a sequence record attributes. Example record attributes include device (e.g., \code{mobile}, \code{desktop}, and \code{tablet}) and source (e.g., \code{paid}, \code{organic}, and \code{direct}).

Our collaborator described a recurring two-stage data analysis workflow: extracting a cohort (e.g., extracting the sessions for South African mobile users) and performing computations on the cohort (e.g., performing funnel analysis on South African mobile sessions). In the following, we discuss the detailed analytic operations employed by our collaborator in this workflow. We do not claim that these operations are exhaustive, but they reflect the analytic requirements of our collaborator. We also note that our framework is extensible: new operations for other applications can be implemented as functions and plugged into the framework easily.

Here are the operations for defining cohorts:

\paragraph{Filter by numerical record attributes} This operation selects sequences whose numerical record attribute falls within a specified range. For example, it can select sessions associated with users between 25 and 40 years old.

\paragraph{Filter by categorical record attributes} This operation selects sequences whose categorical attribute matches a specified value. For example, it can select sessions where device = \code{mobile}.

\paragraph{Filter by patterns} It selects sequences that contain a specified pattern. An example pattern is \code{page\_view $\rightarrow$ add\_to\_cart}. Events in this pattern do not need to be contiguous. Filtering with this pattern returns the sessions in which a \code{page\_view} event occurs before an \code{add\_to\_cart} event, regardless of intervening events.

Patterns may include negative conditions. An example is \code{page\_view $\rightarrow$ add\_to\_cart $\rightarrow$ NOT purchase}. Filtering with this pattern returns the sessions in which a \code{page\_view} event occurs before an \code{add\_to\_cart} event. However, there is no \code{purchase} event after the \code{add\_to\_cart} event. These sessions correspond to shoppers who added items to their cart but never completed a purchase.

\paragraph{Segment by patterns} It extracts a segment within each event sequence. For the pattern \code{add\_to\_cart $\rightarrow$ checkout}, this operation first selects sequences where \code{checkout} occurs after \code{add\_to\_cart}. For these sequences, it truncates each to include only the events between the matching \code{add\_to\_cart} and \code{checkout} events. An analyst can then investigate, for example, the common pattern between adding an item to the cart and initiating checkout.

Here are the operations for performing computations on the cohorts:

\paragraph{Aggregate by numerical record attributes} With a set of event sequences, we can then compute aggregate statistics including average, sum, min, and max over a numerical record attribute. For instance, each session is associated with the age of the shopper. We can find the minimum age across sessions.

\paragraph{Aggregate by categorical record attributes} With a set of event sequences, we can find the distribution of a categorical record attribute. For example, each session is associated with a device type. This operation returns the percentage of sessions associated with mobile, desktop, and tablet devices.

\paragraph{Mine frequent patterns} It mines the top-k most frequent sequential patterns along with the percentage of sequences containing each pattern. An example result is that the pattern \code{page\_view $\rightarrow$ add\_to\_cart $\rightarrow$ checkout} occurs in 35\% of the selected sessions. Our framework employs the PrefixSpan algorithm~\cite{Han2001PrefixSpan}, which is commonly used for sequential pattern mining.

\paragraph{Analyze event attribute distribution} This operation calculates the distribution of an event attribute across all events in the selected sequences. For example, it can report the percentages of events corresponding to \code{add\_to\_cart} and \code{checkout}. It may return that page = \code{checkout} appears in 1\% of the events, indicating that 1\% of the page views are the checkout pages.

\paragraph{Analyze funnel} It involves defining a funnel (i.e., a sequence of events) for the analysis. For each defined step, this operation calculates: (1) the conversion rate (the percentage of sequences reaching the step), (2) the average duration between steps, and (3) the average number of intervening events between steps. For example, in Figure~\ref{hover}, 67.3\% of all sessions have an \code{add\_to\_cart} event. The average duration between \code{add\_to\_cart} and checkout is 51 minutes and 46 seconds.

\subsubsection{Pivot-Table Abstraction}
\label{sec:pivot}

The pivot-table abstraction aligns well with our goals to G1) enable an AI agent to perform analysis on event sequences, G2) assemble a natural language response with data claims, and G3) link each data claim to the concrete analytic steps that are not code.

There are three types of operations in a pivot table: Row operations add rows that represent entities for the analysis; column operations perform computations on each entity; filter operations apply additional filtering constraints to each cell in the pivot table.

\begin{figure}[h]
    \centering
    \includegraphics[width=0.4\linewidth]{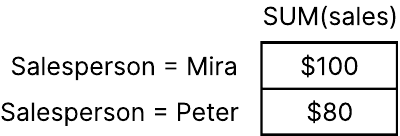}
    \caption{A simple pivot table. A row operation creates one row per salesperson and a column operation creates a column for each salesperson's total sales.}
    \Description{A two-row pivot table showing sales totals by salesperson. The rows are labeled Salesperson equals Mira and Salesperson equals Peter. A single column labeled SUM(sales) contains the values 100 dollars for Mira and 80 dollars for Peter. The table illustrates how a row specifies a subset of records and a column specifies an aggregation performed on those records.}
    \label{pivot}
\end{figure}

A conventional pivot table illustrates the distinction. Suppose a sales dataset contains one record per transaction, including the salesperson and sales amount. A row operation can create one row per salesperson while a column operation can create a column for each salesperson's total sales. In Figure~\ref{pivot}, the first cell indicates that Mira has total sales of \$100.

From this example, the pivot table supports data exploration without writing code (aligns with G1). Each cell in the table represents a finding that can be described in natural language (e.g., Mira's total sales is \$100) (aligns with G2). A cell also represents concrete analytic steps for a finding (aligns with G3). For instance, in Figure~\ref{pivot}, the first cell is obtained by filtering the records for Mira and computing the total sales for these records.

\begin{figure*}[t]
    \centering
    \includegraphics[width=\textwidth]{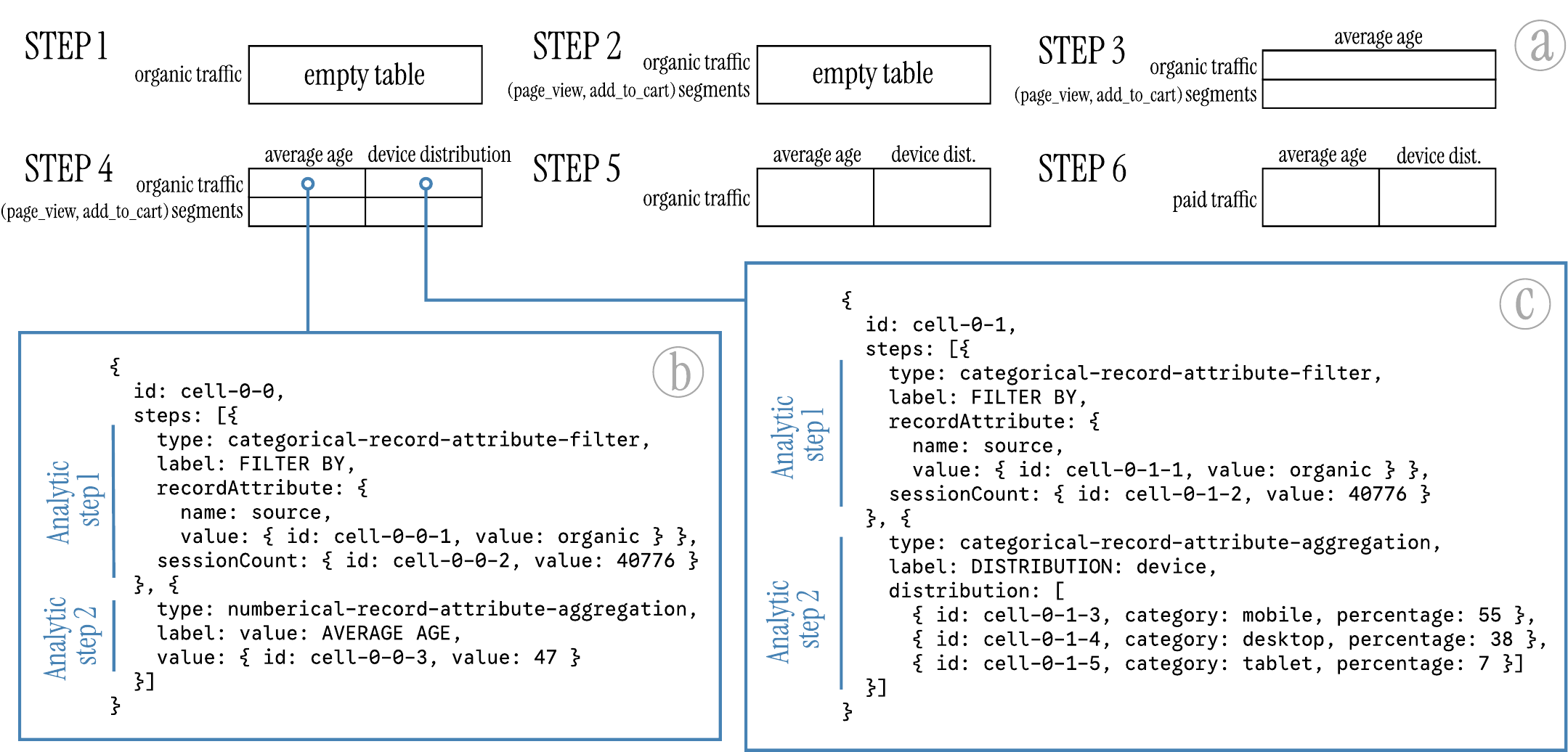}
    \caption{Reference table construction. In (a), step 1 adds a row for organic-traffic sessions; step 2 adds a row for segments between \code{page\_view} and \code{add\_to\_cart}; step 3 adds a column to calculate the average age for both cohorts; step 4 adds a device distribution column; step 5 removes the second row; step 6 changes the first row to paid-traffic sessions. The top-left cell in step 4 represents the average age of users associated with organic-traffic sessions (b). In (b), analytic step 1 filters by source = \code{organic}, yielding 40,776 sessions, while analytic step 2 calculates the average age, yielding 47. The top-right cell in step 4 represents the device distribution for organic traffic (c), where analytic step 1 is filtering and step 2 calculates the percentages of devices.}
    \Description{A three-part diagram illustrating reference table construction. Panel (a) shows six successive table states. Step 1 creates a row for organic-traffic sessions. Step 2 adds a row for segments between page-view and add-to-cart events. Step 3 adds an average-age column, and step 4 adds a device-distribution column, producing a two-row, two-column table. Step 5 removes the second row, and step 6 changes the remaining row to paid traffic. Blue connector lines link two cells in the table at step 4 to the JSON representations below. Panel (b) shows the JSON representation of the average-age cell, containing two analytic steps: filtering for organic traffic, yielding 40,776 sessions, and calculating an average age of 47. Panel (c) shows the JSON representation of the device-distribution cell, containing the same filtering step followed by an aggregation reporting 55 percent mobile, 38 percent desktop, and 7 percent tablet sessions. Both JSON representations contain identifiers for cells and individual results.}
    \label{steps}
\end{figure*}

Besides aligning with our goals, the pivot table abstraction corresponds well with the two types of analytic operations we identified. The four operations for defining cohorts correspond to row operations in the pivot table while the five operations for performing computations correspond to column operations. Filter operations comprise the event sequence analytic operations involving filtering sequences or events, which are the same set as the four operations for defining cohorts. This results in four row operations, five column operations, and four filter operations in our framework.

Figure~\ref{steps} shows how we can use these operations to create a reference table. We first add a row that represents organic-traffic sessions (Figure~\ref{steps}a STEP 1). These sessions are extracted by filtering using a categorical record attribute (source = \code{organic}). At this point, the reference table has rows but no populated cells because no column has been added. Next, we add another row that represents subsequences occurring between \code{page\_view} and \code{add\_to\_cart} (Figure~\ref{steps}a STEP 2). These sessions are extracted by segmenting using the \code{page\_view} and \code{add\_to\_cart} pattern.

We then apply a column operation to find the average age of both cohorts (Figure~\ref{steps}a STEP 3). This operation results in a JSON object representing the analytic steps and results for the organic traffic sessions (Figure~\ref{steps}b). The JSON object shows that analytic step 1 filters sequences where source = \code{organic}, yielding 40,776 sessions. The type of \code{categorical-record-}\code{attribute-} \code{filter} indicates that this step involves filtering by a categorical record attribute. Analytic step 2 calculates the average age, yielding 47. The type of  \code{numerical-} \code{record-}\code{attribute-}\code{aggregation} indicates that this step involves aggregating by a numerical record attribute.

We then apply another column operation to find the device distribution (Figure~\ref{steps}a STEP 4). This operation results in another JSON object in Figure~\ref{steps}c. The JSON object shows that analytic step 1 filters sequences where source = \code{organic} while analytic step 2 calculates that 55\% of sessions used mobile, 38\% used desktop, and 7\% used tablet.

\subsubsection{Tools for Performing Analytic Operations}
\label{sec:tools}

Section~\ref{sec:operations} describes the event sequence analytic operations within our framework. Section~\ref{sec:pivot} illustrates how we organize the analytic operations into row, column, and filter operations to construct a reference table. AI agents interact with the external environment by calling tools. How do we package the row, column, and filter operations into tools?

We provide the agent with a sandbox environment. In this sandbox, the agent can mutate the reference table by calling three tools: the add tool applies row, column, or filter operations to add rows and columns to the reference table or to filter it; the delete tool removes any applied row, column, or filter operation; the update tool modifies an applied operation.

In Figure~\ref{steps}a, an agent constructs a reference table in these steps:

\begin{enumerate}
\item[Step 1:] Invoke the add tool to add a row representing organic-traffic sessions.
\item[Step 2:] Invoke the add tool to add another row representing subsequences between \code{page\_view} and \code{add\_to\_cart}.
\item[Step 3:] Invoke the add tool to add a column calculating the average age for both rows.
\item[Step 4:] Invoke the add tool to add another column calculating the device distribution for both rows.
\item[Step 5:] Invoke the delete tool to remove the row for subsequences between \code{page\_view} and \code{add\_to\_cart}.
\item[Step 6:] Invoke the update tool to change the remaining row from source = \code{organic} to source = \code{paid}.
\end{enumerate}

\subsection{Automatic Insight Generation}
\label{sec:insight}

The above illustrates how an AI agent constructs a reference table. In this reference table, a cell has some analytic steps and the associated data analysis results. From the data analysis results, the agent can assemble a natural language response with data claims. It can link each claim to a cell. When users hover over a claim linked to a cell, the chatbot displays the analytic provenance in that cell.

\begin{figure}[h]
    \centering
    \includegraphics[width=\linewidth]{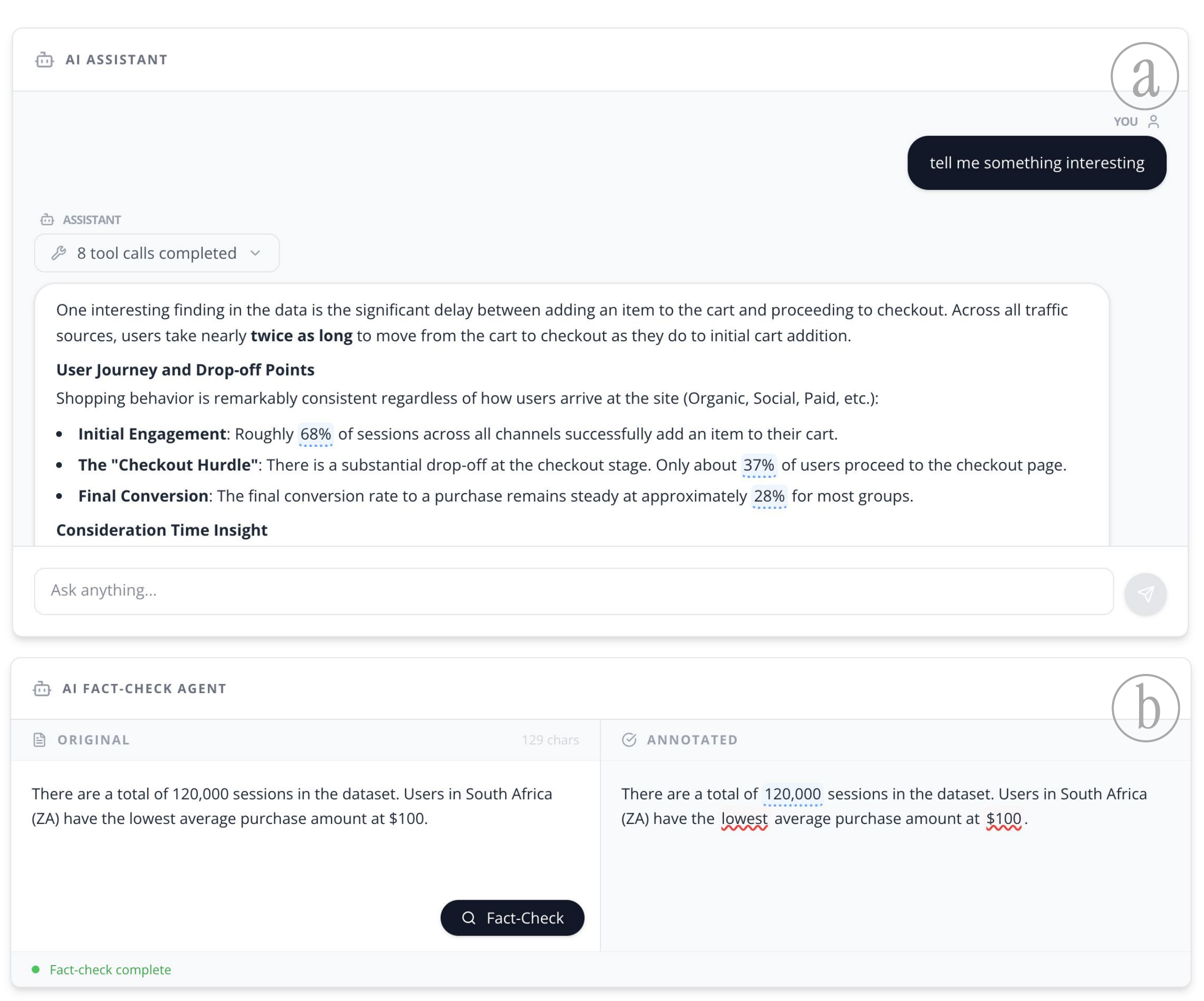}
    \caption{Automated insight generation (a) and post-hoc fact-checking (b). Users can hover over a cited claim and see the analytic steps and results in a floating panel.}
    \Description{Two screenshots of chatbot prototypes arranged vertically. Panel (a) shows the automated insight-generation interface. A user asks the chatbot to provide interesting facts about a dataset, and the chatbot responds with a natural-language summary containing several findings about customer journeys, drop-offs, and conversion behavior. Panel (b) shows the post-hoc fact-checking interface, which presents an original text containing data claims on the left and an annotated version on the right. A Fact-Check button appears beneath the original text, and the interface indicates that fact-checking has completed. A numerical claim in the annotated text is highlighted to indicate a detected inconsistency.}
    \label{prototypes}
\end{figure}

The first use case of this technique is automatic insight generation. We developed a chatbot prototype (Figure~\ref{prototypes}a). Users can ask analytic questions about the underlying event-sequence data and the chatbot responds with cited data claims. Users can hover over a cited claim to visualize the analytic steps that generated the claim.

Below, we discuss how the agent links a claim to the reference table and how the chatbot visualizes the referenced provenance.

\begin{figure*}[t]
    \centering
    \includegraphics[width=\textwidth]{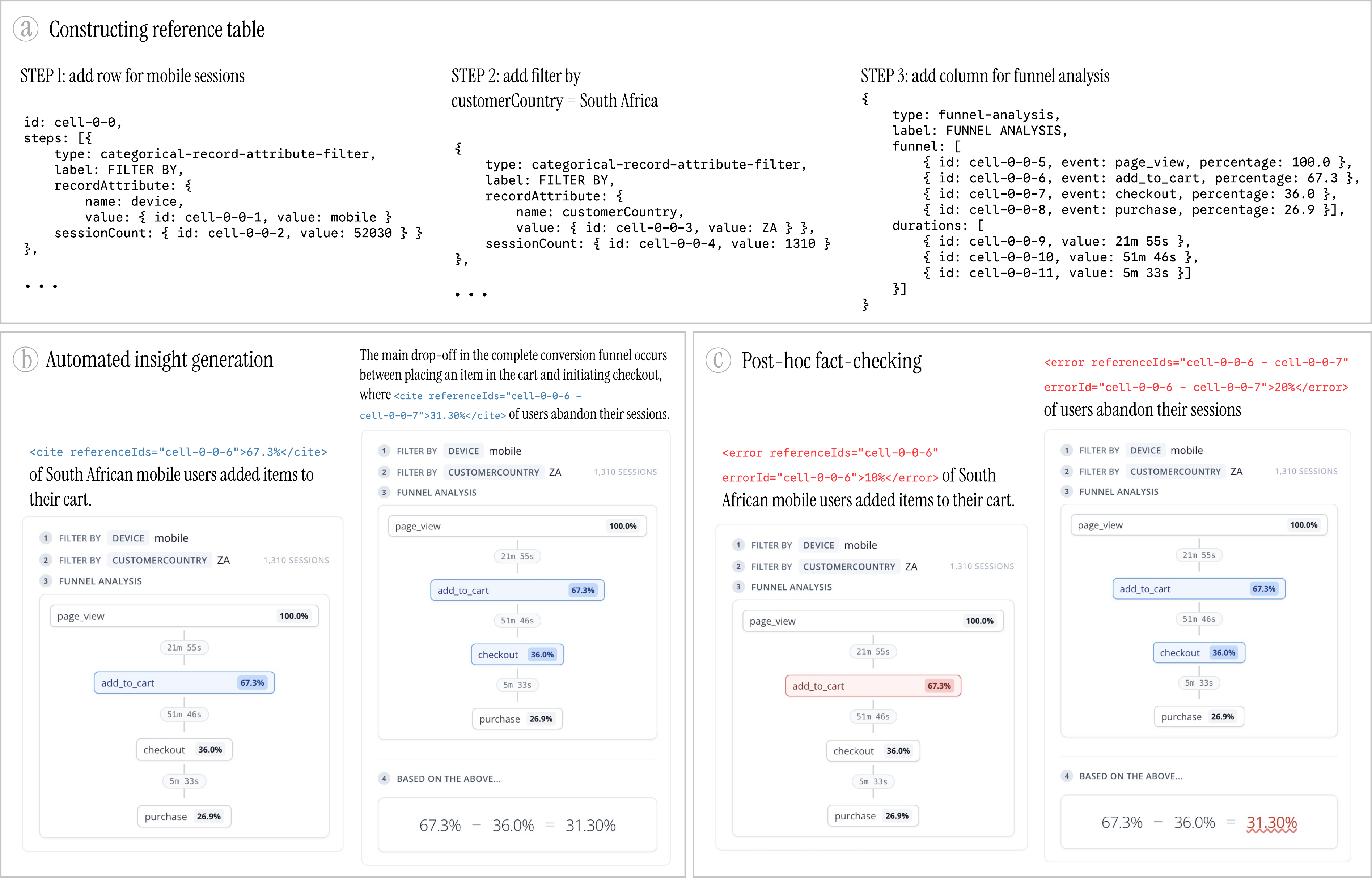}
    \caption{Using a reference table for automated insight generation and post-hoc fact-checking. In (a), a reference table is constructed by adding a row for mobile sessions (step 1), selecting only South African sessions (step 2), and adding a column for funnel analysis (step 3). There is only one cell in the table. The JSON objects in steps 1-3 combined show the complete data in the cell. In (b) left, a user hovers over \textit{``67.3\%''} in \textit{``67.3\% of South African mobile users added items to their cart''} and a funnel with the \code{add\_to\_cart} step highlighted is shown. In (b) right, a user hovers over \textit{``31.3\%''} in \textit{``31.30\% of users abandon their sessions.''} \textit{``31.3\%''} is the percent difference between sessions reaching the \code{add\_to\_cart} step and the \code{checkout} step. These two steps are highlighted and a calculation of the difference is displayed. In (c) left, the \code{add\_to\_cart} step is highlighted in red to show the discrepancy between the \textit{``10\%''} in the claims and the calculated percentage of \textit{``67.3\%''} sessions reaching the \code{add\_to\_cart} step. In (c) right, the difference of \textit{``31.3\%''} is highlighted in red to indicate that the cart abandonment percentage is not the claimed 20\%.}
    \Description{A three-part diagram showing how a reference table supports cited claims and fact-checking. Panel (a) illustrates the construction of a reference-table cell through three operations, each represented by a JSON snippet. The operations select mobile sessions, filter them to South Africa, leaving 1,310 sessions, and perform a four-stage funnel analysis. The funnel results include 100 percent for page view, 67.3 percent for add to cart, 36.0 percent for checkout, and 26.9 percent for purchase. Panel (b) shows two examples of automated insight generation. On the left, a claim citing 67.3 percent displays the funnel with the add-to-cart stage highlighted in blue. On the right, a claim citing 31.30 percent highlights both the add-to-cart and checkout stages and displays their subtraction. Panel (c) shows two examples of post-hoc fact-checking. On the left, an incorrect claim of 10 percent is marked in red, and the funnel highlights the corresponding 67.3 percent result in red. On the right, an incorrect claim of 20 percent is marked in red, and the calculation panel highlights the computed difference of 31.30 percent in red.}
    \label{ref-table}
\end{figure*}

\subsubsection{Citations}

Elements (the cell itself and values) in a cell in the reference table have ids. In the system prompt, we instruct the agent to wrap each data claim in an XML \code{<cite>} tag that includes a \code{referenceIds} attribute. The AI agent references the id of cell elements to link a data claim to the corresponding analytic provenance.

As an example, in Figure~\ref{ref-table}, suppose the user asks, \textit{``Are there any noticeable friction points among our international mobile shoppers?''} and the agent constructs the reference table in Figure~\ref{ref-table}a. The agent then responds \textit{``\code{<cite referenceIds=}\code{"cell-0-0-6">67.3\%</cite>} of South African mobile users added items to their cart.''} (Figure~\ref{ref-table}b left). The id \code{cell-0-0-6} corresponds to the value of 67.3\% sessions that added items to the cart. By referencing \code{cell-0-0-6}, the agent links the claim to the reference table cell that contains the analytic provenance.

The \code{referenceIds} attribute can contain multiple ids linked by arithmetic or logical operators ($+$, $-$, $\times$, $\div$, $>$, $<$, $=$, $\approx$, and $,$). In Figure~\ref{ref-table}b right, the agent further responds that, \textit{``The main drop-off in the complete conversion funnel occurs between placing an item in the cart and initiating checkout, where \code{<cite referenceIds="cell-0-0-6 - cell-0-0-7">31.30\%</cite>} of users abandon their sessions.''} \\\code{cell-0-0-6} corresponds to the percentage of users who added items to their cart while \code{cell-0-0-7} corresponds to the percentage of users who checked out. \code{cell-0-0-6 - cell-0-0-7} represents the difference between the two values.

\begin{figure*}[t]
    \centering
    \includegraphics[width=\textwidth]{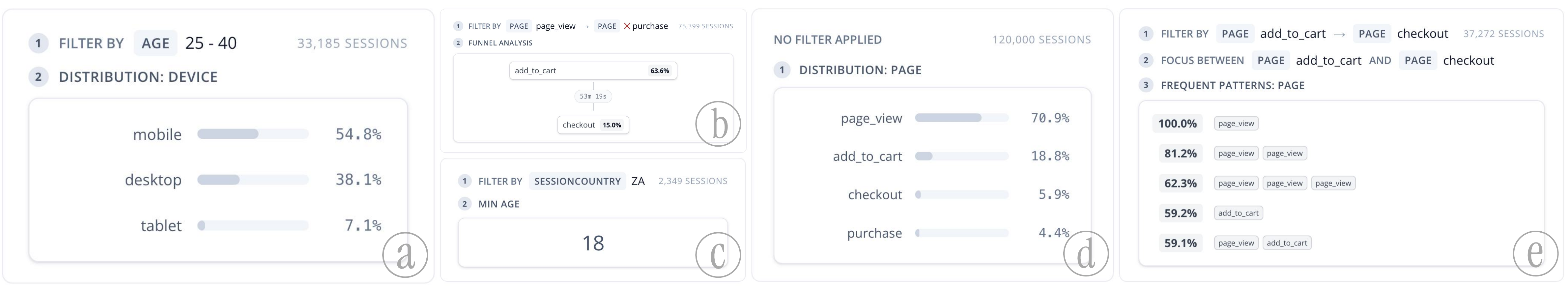}
    \caption{Example visualizations of the nine analytic operations in the floating panel.}
    \Description{Five examples of floating explanation panels, labeled (a) through (e), illustrating nine event-sequence analytic operations. Panel (a) filters sessions to users aged 25 to 40 and displays a horizontal bar chart of device distribution: 54.8 percent mobile, 38.1 percent desktop, and 7.1 percent tablet. Panel (b) filters sequences containing a page-view event without a subsequent purchase and displays a two-stage funnel from add to cart to checkout, including conversion percentages and elapsed time. Panel (c) filters sessions by South Africa and displays a minimum age of 18. Panel (d) displays a horizontal bar chart of event attribute distribution, with page view at 70.9 percent, add to cart at 18.8 percent, checkout at 5.9 percent, and purchase at 4.4 percent. Panel (e) shows three analytic operations: filtering sequences containing add-to-cart followed by checkout, focusing on the events between those two steps, and mining frequent sequential patterns. The final panel lists five patterns and their frequencies, ranging from 100 percent to 59.1 percent. Each example presents numbered analytic steps and their results using compact visual components.}
    \label{vis}
\end{figure*}

\subsubsection{Visualizations}

Within a natural language response, the chatbot highlights cited claims in blue. When users hover over a citation, a floating panel displays the corresponding analytical provenance.

If the agent cites an entire cell, the panel displays all analytical steps and results stored in that cell. If the agent cites a particular element within a cell, the panel displays the complete reference while visually emphasizing that element. For example, in Figure~\ref{ref-table}b left, the claim cites the value associated with the \code{add\_to\_cart} step in the funnel. The floating panel highlights the \code{add\_to\_cart} step.

If a claim cites multiple ids, the panel visualizes each element in order and displays the resulting calculation. For example, in Figure~\ref{ref-table}b right, the panel highlights the \code{add\_to\_cart} event and the \code{checkout} event in the funnel. It further displays the difference between the two percentages at the end.

Figure~\ref{vis} provides example visualizations of the nine analytic operations in Section~\ref{sec:operations} as steps in the floating panel:

\paragraph{Filter by numerical record attributes} As an example, selecting sessions associated with users between 25 and 40 years old is visualized as \raisebox{-0.8ex}{\includegraphics[height=3ex]{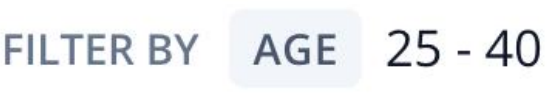}} (Figure~\ref{vis}a step 1).

\paragraph{Filter by categorical record attributes} Selecting sessions from South Africa is visualized as \raisebox{-0.8ex}{\includegraphics[height=3ex]{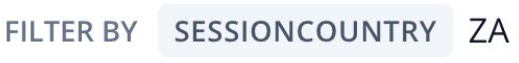}} (Figure~\ref{vis}c step 1).

\paragraph{Filter by patterns} Selecting sessions that have a \code{page\_view} event and no \code{purchase} events afterward is visualized as \newline\raisebox{-0.8ex}{\includegraphics[height=3ex]{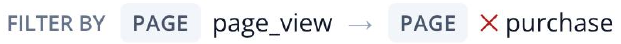}} (Figure~\ref{vis}b step 1).

\paragraph{Segment by patterns} Segmenting by \code{add\_to\_cart $\rightarrow$ checkout} is visualized as first filtering sessions with the pattern \code{add\_to\_cart $\rightarrow$ checkout} \raisebox{-0.8ex}{\includegraphics[height=3ex]{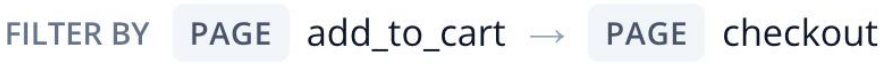}}  (Figure~\ref{vis}b step 1) and second extracting the events between \code{add\_to\_cart} and \code{checkout} \raisebox{-0.8ex}{\includegraphics[height=3ex]{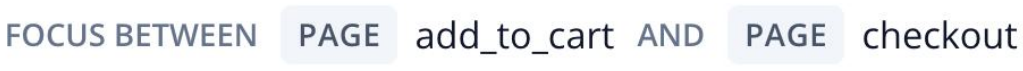}} (Figure~\ref{vis}b step 2).

\paragraph{Aggregate by numerical record attributes} The panel displays an aggregated value directly. Figure~\ref{vis}c step 2 shows the minimum age across all selected sessions.

\paragraph{Aggregate by categorical record attributes} The panel shows a bar graph with the percentages for different record attribute categories. Figure~\ref{vis}a step 2 shows the percentage of devices across sessions.

\paragraph{Mine frequent patterns} The panel shows the patterns together with the percentages of sequences with the pattern. Figure~\ref{vis}e step 3 shows the top-5 frequent patterns. 100\% of sessions have the single-event pattern \code{page\_view}.

\paragraph{Analyze event attribute distribution} The panel displays a bar graph showing the percentages for different event attribute categories. Figure~\ref{vis}d step 1 shows that 70.9\% of events are \code{page\_view} events.

\paragraph{Analyze funnel} The panel displays a funnel showing the percentage reaching each step and the duration or the number of events between steps. Figure~\ref{vis}b step 2 shows a two-step funnel with an \code{add\_to\_cart} event and then a \code{checkout} event.

\subsection{Post-Hoc Fact-Checking}
\label{sec:fact-check}

The second use case of our technique is post-hoc fact-checking. Consumer chatbots such as the Gemini application allow users to double-check factual statements after a response has been generated (Figure~\ref{double-check}). Similarly, once an agent generates event-sequence data claims, we can run the claims through a second agent to fact-check the claims using our technique.

To demonstrate this use case, we developed an automated fact-checking prototype (Figure~\ref{prototypes}b). Users enter event-sequence data claims and click on a button to invoke fact-checking. After the verification, the system highlights correct claims in blue and incorrect ones in red. Users can hover over either type of claim to inspect the underlying analytic steps and results.

\subsubsection{Citations}

The system prompt instructs the AI agent to wrap each correct data claim with a \code{<cite>} tag. It wraps each incorrect data claim with an \code{<error>} tag containing a referenceIds attribute and an errorId attribute. The \code{referenceIds} attribute identifies the references relevant to evaluating the claim. The \code{errorId} attribute identifies the value or relationship that conflicts with the stated claim.

For the statement \textit{``10\% of South African mobile users added items to their cart,''} the agent constructs the table in Figure~\ref{ref-table}a and embeds an \code{<error>} tag around 10\%: \textit{``\code{<error referenceIds="cell-0-0-6" errorId="cell-0-0-6">10\%</error>} of South African mobile users added items to their cart.''} The \code{errorId} points to the calculated value of 67.3\%, making the discrepancy with the stated 10\% explicit.

For the statement \textit{``20\% of users abandon their sessions,''} the agent embeds an \code{<error>} tag around 20\%: \textit{``\code{<error referenceIds=}\code{"cell-0-0-6 - cell-0-0-7" errorId="cell-0-0-6 - cell-0-0-7">20\%</error>} of users abandon their sessions.''} The \code{errorId} indicates a discrepancy between 20\% and the calculated 67.3\% - 36\% = 31.3\%.

\subsubsection{Visualizations}

The interface highlights correct claims in blue and incorrect claims in red. Hovering over a claim opens the same floating panel used for automatic insight generation.

In the floating panel, the element referenced by \code{errorId} is highlighted in red to make the discrepancy between the claim and the data salient. For \textit{``\code{<error referenceIds="cell-0-0-6"} \code{errorId=} \code{"cell-0-0-6">10\%</error>} of South African mobile users added items to their cart,''} \code{cell-0-0-6} corresponds to the \code{add\_to\_cart} step in the funnel and we highlight this step in red (Figure~\ref{ref-table}c left). For \textit{``\code{<error referenceIds="cell-0-0-6 - cell-0-0-7" errorId="cell-0-0-6 - cell-0-0-7">20\%</error>} of users abandon their sessions,''} \code{cell-0-0-6 - cell-0-0-7} corresponds to the difference, and we highlight the difference in red (Figure~\ref{ref-table}c right).

\section{Crowdsourced Experiment 1}

One use case of our technique is automated insight generation: given some data, an agent analyzes the data, synthesizes a natural language response, and cites the event-sequence data claims in the response. The rapid advancement of LLMs suggests a future where the reliability of AI-generated analyses is significantly higher. This raises a critical question: if all data claims are perfectly accurate, how do the citations shape users' assessments of the claims?

To investigate this, we designed a between-subjects experiment in which participants reviewed a series of correct event-sequence data claims under one of three explanation conditions: citations, code snippets, or no explanations (baseline). We measured perceived accuracy of insights, perceived overall system reliability, and system understanding.

\subsection{Methodology}

\subsubsection{Fact Preparation}

We used the same Kaggle e-commerce dataset described in Section~\ref{sec:operations}~\cite{ElHusseini2024EcommerceClickstream}. We chose this dataset because its domain (i.e., online shopping behavior) is accessible to laypeople without specialized training.

To prepare the citation condition, we loaded this dataset into our chatbot prototype (Figure~\ref{prototypes}a) and prompted it using the query ``I want to get a sense for the data. Tell me some interesting facts about the data.'' We aimed to simulate a typical chatbot interaction where a user hovers over one citation, views the analytic steps, and forms an assessment of the claim. Hence, we designed each task to present a single statement with one citation and one explanation panel. From the natural language response from our chatbot prototype, we collected ten statements each with exactly one cited data claim.

With the 10 statements, we prepared the code condition next. We supplied Claude Code with the Kaggle dataset and instructed it to write Python code snippets that would produce the corresponding data claims. During the prompting, we followed the common practice where one snippet can generate multiple data claims. As a result, Claude Code produced four code snippets that collectively generated the 10 claims. All snippets include inline comments to improve comprehension. Since chatbots like the Gemini application display console output for each code snippet, we executed each snippet and collected the console output. In the end, we obtained four code snippets and four console outputs for the 10 data claims.

\begin{figure*}[t]
    \centering
    \includegraphics[width=\textwidth]{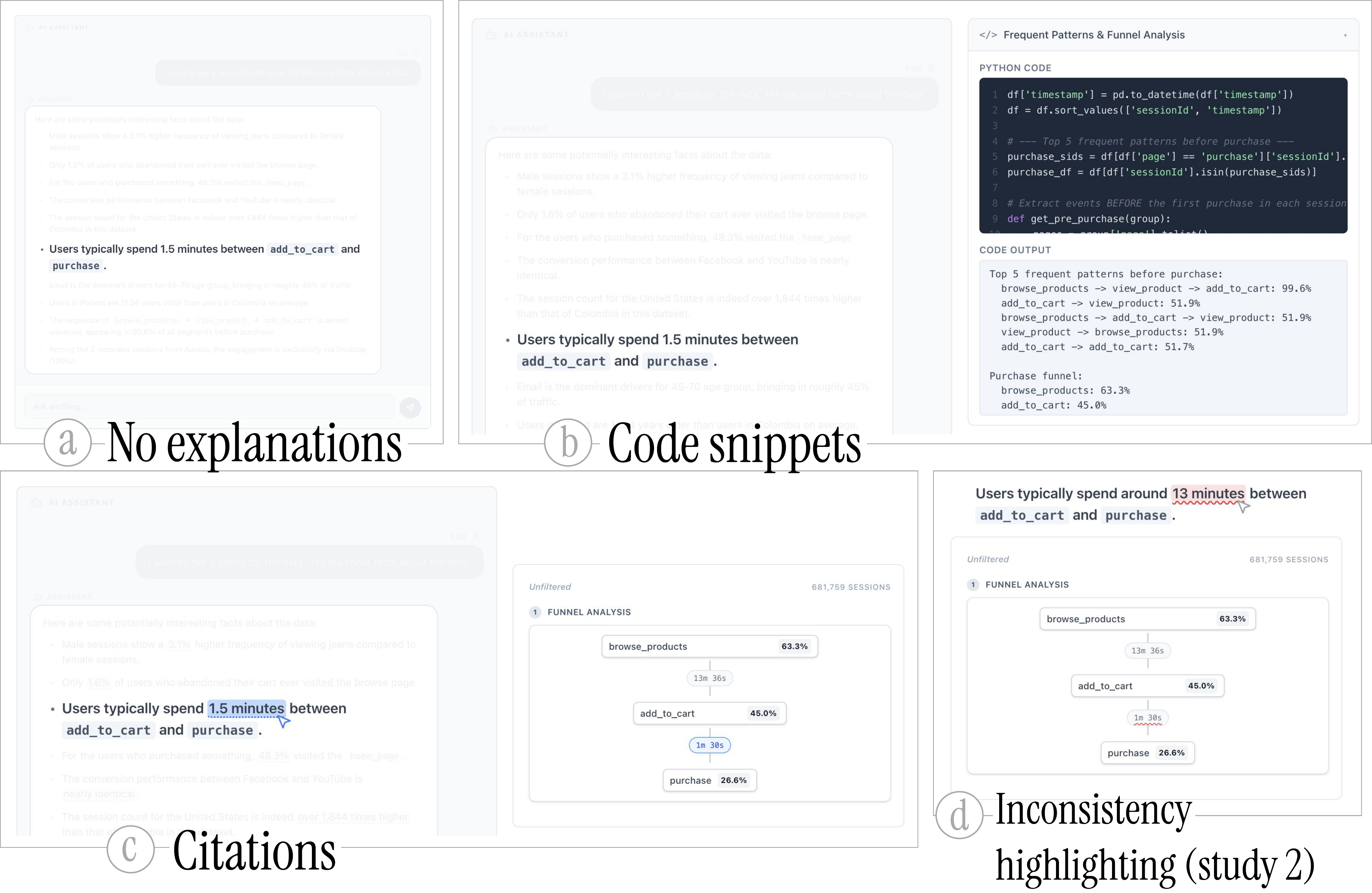}
    \caption{Experimental conditions for experiments 1 and 2. In the no explanation condition (a), the interface for each task shows all 10 statements. Only the statement being rated is highlighted while the rest are grayed out. In the code condition (b), participants saw a statement highlighted on the left and code snippets presented on the right. In the citation condition (c), participants saw a statement on the left and a floating panel with explanatory analytic steps and results on the right. In Experiment 2, for incorrect statements, the data claim is highlighted in red on the left and the floating panel highlights the inconsistency on the right (d).}
    \Description{Four screenshots showing the experimental interfaces. Panel (a), No explanations, displays a chatbot response containing several data statements, with the statement being evaluated emphasized and the surrounding text faded. Panel (b), Code snippets, displays the same type of highlighted statement on the left and a Python code snippet with its console output on the right. Panel (c), Citations, displays the highlighted statement on the left and a floating explanation panel on the right, showing a conversion funnel and the durations between stages. Panel (d), Inconsistency highlighting, shows an incorrect statement reporting 13 minutes between add-to-cart and purchase events. The incorrect duration is underlined in red, and the explanation panel highlights the corresponding discrepancy in the funnel in red.}
    \label{conditions}
\end{figure*}

\subsubsection{Conditions}

We compared three explanation types: no explanations, code snippets and citations.

\paragraph{No explanations (Figure~\ref{conditions}a)} Participants reviewed ten statements one by one with no supporting information. This condition served as the baseline. For each task, we asked participants to rate the perceived accuracy of a statement. In the task interface, we presented all statements. The statement being rated was highlighted while the other nine were grayed out.

\paragraph{Code snippets (Figure~\ref{conditions}b)} Chatbots like the Gemini application provide code snippets together with the natural language response even when users may not be proficient in programming. In this condition, participants reviewed the same ten statements with four Python code snippets displayed alongside. In the task interface, we presented all statements on the left with the currently rated one being highlighted. We presented all four code snippets and the respective console outputs on the right.

\paragraph{Citations (Figure~\ref{conditions}c)} Participants reviewed the same ten statements with one cited data claim each. In the task interface, a cursor pointed to the cited data claim in the statement being rated. We displayed the corresponding analytic steps and results in a floating panel on the right.

\subsubsection{Participants}

A power analysis indicated that detecting a medium effect size ($f = 0.25$) with a one-way ANOVA at $\alpha = 0.05$ and power of 0.8 required 90 participants (30 per condition). 

We recruited participants from Prolific with the following inclusion criteria: an approval rating of 95\% or above, English proficiency, a work function related to data analysis, and prior experience using AI chatbots. We did not require programming experience because we targeted the user base of consumer chatbots such as Gemini and ChatGPT, which generate data claims with code snippets regardless of users' coding ability. To screen for attentiveness, we asked 4 comprehension questions about event sequence analysis before the main task; participants who answered fewer than 2 correctly were excluded. They were compensated \pounds13.59 per hour on average.

Participants (61 male, 29 female) were aged 18--24 (15), 25--34 (43), 35--44 (15), 45--54 (13), and 55--64 (4). For data analysis expertise, 24 reported beginner-level, 34 intermediate, 29 advanced, and 3 expert experience. For chatbot usage, 42 used chatbots daily, 31 often, 16 sometimes, and 1 rarely. Among participants in the code condition, 3 had no coding experience, 14 beginner, 8 intermediate, and 5 advanced.

\subsubsection{Procedure}

We randomly assigned participants to one of the three conditions. Participants first completed a demographic survey and the screening questions. They then completed a practice task to familiarize themselves with the interface. The practice task mirrored the main task but used different statements.

In the main task, participants reviewed ten statements with data claims. They rated one statement at a time. The presentation order of the 10 statements was randomized to avoid order effects. For each statement, we asked, \textit{``Based on the above information, how accurate or inaccurate does the above statement seem to you?''} Participants reported their perceived accuracy on a 7-point Likert scale.

After the main task, participants answered two post-task questions. First, they rated the overall system reliability on a 7-point Likert scale and provided reasons for their rating. Second, they rated their understanding about the system on a 7-point scale and detailed their justifications in an open-ended question. The study took approximately 23 minutes on average.

\subsubsection{Measures}

We collected three measures:

\paragraph{Perceived accuracy} For each statement, participants rated \textit{``Based on the above information, how accurate or inaccurate does the above statement seem to you?''} on a 7-point Likert scale ($-3$: very inaccurate to $+3$: very accurate). For each participant, we computed the average rating.

\paragraph{Perceived overall reliability} After participants rated the ten statements, we asked, \textit{``Overall, how reliable do you think the facts generated by this chatbot are?''} on a 7-point Likert scale ($-3$: very unreliable to $+3$: very reliable).

\paragraph{System understanding} Besides perceived overall reliability, we further asked, \textit{``How well do you understand the steps the AI agent took when generating the insights?''} on a 7-point scale ($-3$: did not understand to $+3$: understand completely).

For each quantitative measure, we conducted Kruskal-Wallis tests to examine the main effect of explanation type. When a significant main effect was found, we conducted post-hoc Wilcoxon rank-sum tests with Holm-Bonferroni correction for pairwise comparisons.

For the two post-task questions regarding perceived overall reliability and system understanding, we also asked participants to explain their ratings in open-ended questions. We conducted a thematic analysis~\cite{AHMED2025100198} of these qualitative responses to complement the quantitative findings.

\subsubsection{Hypotheses}

Schemmer et al.~\cite{Schemmer2023AoR} showed that explanations tend to increase reliance on AI advice. While their study focused on AI advice and not data analysis, we posited that the finding would carry over to event-sequence data claims.

\paragraph{H1.1} Participants with explanations (citations or code) will report higher perceived accuracy than those without explanations.

\paragraph{H1.2} Participants with explanations (citations or code) will report higher perceived overall reliability than those without explanations.

Xie et al.~\cite{WaitGPT2024} found that structured visual representations of analytic steps are more comprehensible than raw code. Participants in the citation condition reviewed analytic steps visualized in a floating panel. We expected them to be able to understand the explanations of data claims better than participants in the code condition and the baseline.

\paragraph{H1.3} Participants with citations will report better system understanding than those with code or no explanations.

\begin{figure*}[t]
    \centering
    \includegraphics[width=\textwidth]{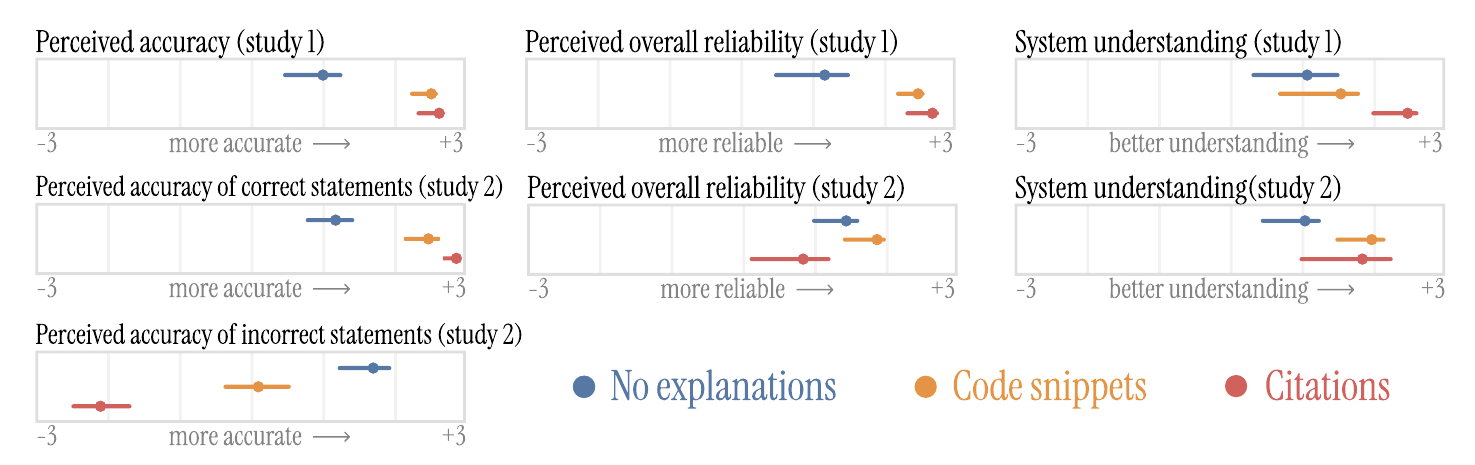}
    \caption{Quantitative results from crowdsourced experiments 1 and 2. Error bars show 95\% bootstrapped confidence intervals.}
    \Description{Seven horizontal dot-and-error-bar plots arranged in three columns, comparing the no-explanation, code-snippet, and citation conditions across two experiments. Blue represents no explanations, orange represents code snippets, and red represents citations. Each plot has a horizontal rating scale from minus 3 to plus 3, and error bars represent 95 percent bootstrapped confidence intervals. The top row shows Experiment 1: perceived accuracy, perceived overall reliability, and system understanding. Both code and citations receive higher accuracy and reliability ratings than no explanations, while citations receive the highest understanding rating. The middle row shows Experiment 2: perceived accuracy of correct statements, overall reliability, and system understanding. Citations receive the highest accuracy rating for correct statements, code receives the highest overall reliability rating, and code and citations receive similar understanding ratings. The bottom-left plot shows perceived accuracy of incorrect statements in Experiment 2. Citations receive the lowest mean rating, approximately minus 2.17, code receives a rating near zero, and no explanations receives the highest rating, approximately 1.63.}
    \label{crowd}
\end{figure*}

\subsection{Results}

Figure~\ref{crowd} shows the quantitative results for the two crowdsourced experiments. We observed that explanations (code or citations) tended to increase perceived accuracy and perceived reliability compared to no explanations. Also, citations led to significantly better system understanding than code and no explanations.

In the following, we provide the detailed analysis.

\subsubsection{Perceived Accuracy (H1.1 Supported)}

Participants in the citation condition rated the statements the most accurate (\textit{M}=2.55, \textit{SD}=0.53), followed by the code condition (\textit{M}=2.44, \textit{SD}=0.54), and the no-explanation baseline (\textit{M}=0.93, \textit{SD}=1.16). A Kruskal-Wallis test indicated a significant main effect of explanation type on perceived accuracy ($\chi^2(2)=40.80$, \textit{p}<.001). Post-hoc pairwise comparisons showed that both the citation condition (\textit{p}<.001) and the code condition (\textit{p}<.001) had significantly higher perceived accuracy than the baseline. The citation and code conditions did not differ significantly (\textit{p}=.33).

\subsubsection{Perceived Overall Reliability (H1.2 Supported)}

Participants in the citation condition reported the highest perceived reliability (\textit{M}=2.60, \textit{SD}=0.62), followed by the code condition (\textit{M}=2.40, \textit{SD}=0.56), and the no-explanation condition (\textit{M}=1.10, \textit{SD}=1.45). Explanation type had a significant main effect ($\chi^2(2)=27.27$, \textit{p}<.001). Both the citation condition (\textit{p}<.001) and the code condition (\textit{p}<.001) had significantly higher perceived reliability than the no-explanation condition. The citation and code conditions did not differ significantly (\textit{p}=.12).

\subsubsection{System Understanding (H1.3 Supported)}

Participants in the citation condition reported the highest system understanding (\textit{M}=2.40, \textit{SD}=0.86), followed by the code condition (\textit{M}=1.47, \textit{SD}=1.74), and the no-explanation condition (\textit{M}=1.00, \textit{SD}=1.72). Explanation type had a significant main effect ($\chi^2(2)=13.99$, \textit{p}<.001). The citation condition had significantly higher system understanding than both the code condition (\textit{p}=.033) and the no-explanation condition (\textit{p}<.001). The code and no-explanation conditions did not differ significantly (\textit{p}=.22).

\section{Crowdsourced Experiment 2}

Another use case of our technique is post-hoc fact-checking. During post-hoc fact-checking, an agent first writes code to analyze the data and synthesize a natural language response. Similar to the workflow to double-check a claim in Figure~\ref{double-check}, users then trigger fact-checking, and the agent leverages our technique to verify and cite the data claims within the response. For correct data claims, users see the supporting data highlighted in blue in the floating panel (as in Figure~\ref{ref-table}b left) while for incorrect data claims, users see the inconsistency highlighted in red in the panel (as in Figure~\ref{ref-table}c left).

Experiment 1 shows that for correct data claims, citations tend to improve perceived accuracy when compared with no explanations. For incorrect data claims, how do the citations with inconsistency highlighting affect users' assessments of the claims?

To investigate this, we conducted a second between-subjects experiment. As in the first experiment, participants reviewed a series of statements under the three explanation conditions. However, four of the ten statements contained errors.

\subsection{Methodology}

\subsubsection{Fact Preparation}

We started with the same ten statements from Experiment 1. Based on research on common failure modes of AI-generated data analyses, we introduced errors in four of them. Common errors include misinterpretation of computed values~\cite{zhou2025deepjsoneval} and omission of analytic steps such as filters~\cite{WaitGPT2024}.

For misinterpretation errors, the code produces a correct result but the AI agent generates an incorrect data claim. For instance, the code correctly computes that the session count for the United States is 1,844 times higher than that of Colombia. However, the agent rounds 1,844 to 2,000 and wrongly states that ``The session count for the United States is indeed over 2,000 times higher than that of Colombia in this dataset.''

For missing-filter errors, the code misses a filtering step, results in an incorrect calculation, and hence, an incorrect data claim. For example, an agent attempts to compute the source distribution for the 45–70 age group. It omits the age filter and computes the source distribution across all ages instead. It claims that \textit{``45\% of traffic for the 45–70 age group is from email''} when the correct statement is \textit{``45\% of traffic for all age groups is from email.''}

For the citation condition, we did not change the six correct statements. For the two incorrect statements with interpretation errors, we updated the statement text. For example, we changed \textit{``the session count for the United States is indeed over 1,844 times higher than that of Colombia in this dataset''} to \textit{``the session count for the United States is indeed over 2,000 times higher than that of Colombia in this dataset.''} For the two incorrect statements with missing filters, we removed a filter from the floating panel. For the statement \textit{``email is the dominant driver for the 45–70 age group, bringing in roughly 45\% of traffic,''} we removed the step that selects sessions from users aged 45–70. We highlight both the data claim in the incorrect statement and the inconsistency in the floating panel in red (Figure~\ref{conditions}d).

For the code condition, we again did not change the six correct statements. For the two incorrect statements with interpretation errors, we updated the statement text without changing the code logic. For the two incorrect statements with missing filters, we removed a filtering step in the code.

\subsubsection{Conditions, Participants, and Procedure}

The three conditions and the experimental procedure were identical to Experiment~1. We recruited 90 new participants from Prolific (30 per condition) using the same inclusion criteria. Participants were aged 18--24 (10), 25--34 (47), 35--44 (21), 45--54 (7), and 55--64 (5). 48 identified as male and 42 as female. Regarding data analysis expertise, 26 reported beginner-level experience, 33 intermediate, 23 advanced, and 8 expert. For chatbot usage, 52 used chatbots daily, 27 often, 6 sometimes, and 5 rarely. Among participants in the code condition, 5 had no coding experience, 11 beginner, 11 intermediate, 1 advanced, and 2 expert. Participants were compensated \pounds13.02 per hour on average. The study took approximately 24 minutes on average.

\subsubsection{Measures}

We collected four measures. Perceived overall reliability and system understanding were the same as in Experiment 1. For perceived accuracy, we calculated two measures:

\paragraph{Perceived accuracy of incorrect statements} For each statement, participants rated \textit{``Based on the above information, how accurate or inaccurate does the above statement seem to you?''} on a 7-point scale. For each participant, we computed the average rating over the four incorrect statements.

\paragraph{Perceived accuracy of correct statements} For each participant, we computed the average perceived accuracy rating over the six correct statements.

We measured the perceived accuracy of incorrect statements since we were interested in whether citations with inconsistency highlighting would reduce the perceived accuracy of erroneous statements. We measured the perceived accuracy of correct statements since we wanted to confirm the findings in Experiment 1: citations improve the perceived accuracy of correct statements when compared with no explanations.

\subsubsection{Hypotheses}

In Experiment 1, all statements were correct. Participants tended to perceive the statements to be more accurate when presented with citations. We expected to have a similar observation for the correct statements in this experiment.

\paragraph{H2.1} Participants with explanations (citations or code) will report higher perceived accuracy for correct statements than those without explanations.

Kim et al.~\cite{Kim2025Fostering} showed that inconsistency highlighting in explanations for factual claims can reduce overreliance. Hence, we formulated these hypotheses:

\paragraph{H2.2} Participants with citations will report lower perceived accuracy on incorrect statements than those with code or no explanations, because inconsistency highlighting makes errors more salient.

\paragraph{H2.3} Participants with explanations (citations or code) will report lower perceived overall reliability than those without explanations.

\paragraph{H2.4} Following H1.3, participants with citations will report better system understanding than those with code or no explanations.

\subsection{Results}

From Figure~\ref{crowd}, we again observed that for accurate statements, explanations (code or citations) tended to increase perceived accuracy compared with no explanations. Also, for incorrect statements, citations appear to reduce perceived accuracy when compared with both code and no explanations. Participants also tended to perceive the overall reliability to be lower and have a better system understanding with citations than with no explanations.

In the following, we provide the detailed analysis.

\subsubsection{Perceived Accuracy of Correct Statements (H2.1 Supported)}

Participants in the citation condition rated the correct statements the most accurate (\textit{M}=2.79, \textit{SD}=0.29), followed by the code condition (\textit{M}=2.40, \textit{SD}=0.72), and the no-explanation baseline (\textit{M}=1.11, \textit{SD}=0.95). A Kruskal-Wallis test indicated a significant main effect of explanation type on perceived accuracy of correct statements ($\chi^2(2)=40.56$, \textit{p}<.001). Post-hoc pairwise comparisons showed that both the citation condition (\textit{p}<.001) and the code condition (\textit{p}<.001) had significantly higher perceived accuracy than the baseline. The citation and code conditions did not differ significantly (\textit{p}=.12).

\subsubsection{Perceived Accuracy of Incorrect Statements (H2.2 Supported)}

Participants in the citation condition rated the incorrect statements the least accurate (\textit{M}=$-$2.17, \textit{SD}=1.15), followed by the code condition (\textit{M}=0.03, \textit{SD}=1.29), and the no-explanation condition (\textit{M}=1.63, \textit{SD}=1.05). A Kruskal-Wallis test indicated a significant main effect of explanation type on perceived accuracy ($\chi^2(2)=57.17$, \textit{p}<.001). The citation condition had significantly lower perceived accuracy than both the code condition (\textit{p}<.001) and the no-explanation condition (\textit{p}<.001). The code condition also had significantly lower perceived accuracy than the no-explanation condition (\textit{p}<.001). Inconsistency highlighting in citations appears to be effective at surfacing errors, while code snippets and console outputs might require closer inspection to spot errors.

\subsubsection{Perceived Overall Reliability (H2.3 Partially Supported)}

A Kruskal-Wallis test revealed a significant main effect of explanation type on perceived overall reliability ($\chi^2(2)=8.74$, \textit{p}=.013). Post-hoc pairwise comparisons showed that participants in the citation condition (\textit{M}=0.77, \textit{SD}=1.59) rated reliability significantly lower than those in the code condition (\textit{M}=1.80, \textit{SD}=0.81; \textit{p}=.014), but not significantly lower than those without explanations (\textit{M}=1.37, \textit{SD}=0.96; \textit{p}=.20). The difference between code and no explanation was also not significant (\textit{p}=.13).

While Experiment~1 suggested that explanations improve perceived reliability, the open-ended responses indicated that error salience might offset the improvements. The offsetting effects may explain why neither explanation condition differed significantly from no explanations. A participant in the citation condition noted, ``\textit{The bot exhibits inconsistencies in basic calculations, rendering the analysis unreliable.}''

A stronger error salience with citations than with code might explain the lower perceived reliability with citations. Citations directly highlighted inconsistencies between stated facts and the underlying data, whereas code required participants to trace the logic themselves---and some could not detect the errors at all: ``\textit{The facts mostly matched the code outputs and calculations, so the insights appear accurate.}''

\subsubsection{System Understanding (H2.4 Partially Supported)}

Participants in the code condition reported the highest system understanding (\textit{M}=1.90, \textit{SD}=0.99), followed by the citation condition (\textit{M}=1.77, \textit{SD}=1.79), and the baseline (\textit{M}=0.97, \textit{SD}=1.40). A Kruskal-Wallis test indicated a significant main effect of explanation type on system understanding ($\chi^2(2)=9.93$, \textit{p}=.007). Both the citation condition (\textit{p}=.022) and the code condition (\textit{p}=.022) had significantly higher system understanding than the no-explanation condition. However, the citation and code conditions did not differ significantly (\textit{p}=.32). This contrasts with Experiment~1, where citations led to significantly higher understanding than code. Experiment~1 suggested that citations improve system understanding, likely because they make the analytic steps more accessible. With obvious errors, some participants in the citation condition were confused about how the system derives its answers: ``\textit{The steps themselves were very understandable on the right. It is therefore somewhat incomprehensible why some major errors occurred. I'm not sure what the AI did here.}'' This erosion of confidence may have narrowed the gap between the citation and code conditions observed in Experiment~1.

\section{Discussion}

Here, we discuss how to extend our technique to support more analytic operations, the implications of our crowdsourced experiments for appropriate reliance, and the study limitations.

\subsection{Technique Extensibility}
\label{sec:extend}

While Section~\ref{sec:operations} focuses on the analytic operations in the application of clickstream analysis, our technique is highly extensible. It can be expanded to support a broader set of event-sequence analytic operations by following the procedure outlined in Section~\ref{sec:table}. First, we curate a new set of operations for the application (Sec.~\ref{sec:operations}). Second, we classify the analytic operations into row, column, and filter operations (Sec.~\ref{sec:pivot}): operations for defining cohorts are row operations; operations for performing computations on cohorts are column operations; operations involving filtering of sequences or events are filter operations. Third, we implement the new operations as functions that can be invoked by the add, delete, and update tools (Sec.~\ref{sec:tools}).

As an example, for outcome analysis, we may identify edge analysis~\cite{Gotz2014DecisionFlow} as a new analytic operation (step 1). Since this operation involves performing computations on a cohort, we classify this operation as a column operation (step 2). Finally, we can implement edge analysis as a function that takes a set of sequences as the input to calculate the odds ratio of each event attribute with respect to a specified outcome event as the output (step 3).

\subsection{Citations and Appropriate Reliance}

When evaluating an explanation technique, researchers often assess whether the explanation fosters appropriate reliance: users should trust the AI outputs when they are correct and reject them otherwise. Crowdsourced experiment 1 showed that participants tended to perceive correct data claims to be more accurate with citations when compared with no explanations. Crowdsourced experiment 2 indicated that participants appeared to perceive incorrect data claims to be less accurate with citations when compared with code and no explanations. Increasing perceived accuracy of correct claims while reducing perceived accuracy of incorrect ones is the premise of appropriate reliance.

Our crowdsourced experiments also echoed studies in domains other than data analysis. For example, Schemmer et al.~\cite{Schemmer2023AoR} reported that explanations tend to increase reliance on AI advice. Kim et al.~\cite{Kim2025Fostering} suggested that inconsistencies in explanations for factual claims can reduce overreliance. Different from this body of work, we focused on studying a different domain of event sequence analysis. We further contributed a comparison between citations and code as an explanation mechanism: for correct data claims, we do not have evidence that citations improve the perception of accuracy compared with code; for incorrect data claims, we found that citations reduced the perception of accuracy compared with code, presumably because the analytic steps and results in the floating window are more accessible than code.

\subsection{Study Limitations}

We conducted two crowdsourced experiments to examine how citations might shape users' assessment of event-sequence data claims. We observed that citations could foster appropriate reliance. However, we are prudent in drawing conclusions about the practical significance of this finding. Our crowdsourced experiments measured perceived accuracy and perceived reliability---not whether participants made better decisions. While perceived accuracy is a commonly used proxy~\cite{Zhang2020Effect, Yin2019AccuracyTrust}, a user who perceives a correct data claim as accurate may still fail to act on it, and a user who perceives an incorrect claim as inaccurate may not know how to correct it. Future work should investigate whether citations lead to better decision outcomes in more realistic analytic tasks.

\section{Conclusion}

In this paper, we introduced a technique linking event-sequence data claims to their analytic provenance in chatbots. To develop this technique, we identified a set of analytic operations for clickstream analysis, organized them through a pivot-table abstraction, and developed tools so an AI agent can construct a reference table for citing data claims within natural language responses. We demonstrated the technique in two applications: automated insight generation and post-hoc fact-checking. In both applications, users can inspect how a claim was derived by hovering over it rather than by reading code. Two crowdsourced experiments further suggested that citations could help foster appropriate reliance: they tended to raise the perceived accuracy of correct claims relative to no explanation and, with inconsistency highlighting, lowered the perceived accuracy of incorrect claims relative to both code and no explanation. These findings suggest that linking individual data claims to inspectable analytical steps and results can help users form assessments that are more sensitive to whether the available evidence supports or contradicts an AI-generated analysis.

\bibliographystyle{ACM-Reference-Format}
\bibliography{reference}

\end{document}